\documentclass[letterpaper, 10 pt, conference]{ieeeconf}  

\IEEEoverridecommandlockouts                              

\usepackage{graphics} 
\usepackage{epsfig} 
\usepackage{times} 
\usepackage{amsmath}
\usepackage{amsfonts}
\usepackage{amssymb}
\usepackage[usenames,dvipsnames]{xcolor}
\usepackage{multirow}
\usepackage{graphicx}
\usepackage{hyperref}
\usepackage{booktabs}
\usepackage{comment}
\usepackage{soul}
\usepackage{float}
\usepackage{subcaption}
\usepackage{bm}

\newcommand{\re}[1]{{\color{black}#1}}
\newcommand{\ree}[1]{{\color{black}#1}}
\usepackage{algorithm}
\usepackage{algorithmic}

\title{\LARGE \bf
GuideLight: ``Industrial Solutions'' Guidance  for More Practical Traffic Signal Control Agents
}

\author{Haoyuan Jiang$^{1}$$^{\dagger}$, Xuantang Xiong$^{2}$$^{\dagger}$, Ziyue Li$^{3}$$^{\dagger}$$^{\ast}$, Hangyu Mao$^{4}$, Guanghu Sui$^{4}$, \\ Jingqing Ruan$^{2}$, Yuheng Cheng$^{2}$, Hua Wei$^{5}$, Wolfgang Ketter$^{3}$, Rui Zhao$^{4}$
\thanks{$^{1}$Haoyuan Jiang was a researcher at SenseTime, this work was done there, now in Baidu Inc. {\tt\small jianghaoyuan@zju.edu.cn}}%
\thanks{$^{2}$Xuantang Xiong, Jingqing Ruan, and Yuheng Cheng were research interns at SenseTime Research. They are PhD candidates at the Chinese Academy of Sciences and the Chinese University of Hong Kong.} 
\thanks{$^{3}$Ziyue Li and Wolfgang Ketter are with the Information Systems Department at the University of Cologne, Germany. {\tt\small zlibn@wiso.uni-koeln.de}}%
\thanks{$^{4}$Hangyu Mao, Guanghu Sui, and Rui Zhao are with SenseTime Research. {\tt\small maohangyu@sensetime.com}, {\tt\small zhaorui@sensetime.com}}; 
\thanks{$^{5}$Hua Wei is an assistant professor in the School of Computing and Augmented Intelligence at Arizona State University. {\tt\small hua.wei@asu.edu}}%
\thanks{$\dagger$ These authors contribute equally to this work.} 
\thanks{$\ast$ The corresponding author.} %
}

\begin{document}

\maketitle
\thispagestyle{empty}
\pagestyle{empty}

\begin{abstract}
Currently, traffic signal control (TSC) methods based on reinforcement learning (RL) have proven superior to traditional methods. However, most RL methods face difficulties when applied in the real world due to three factors: input, output, and the cycle-flow relation. \re{The industry's observable input is much more limited than simulation-based RL methods. For real-world solutions, only flow can be reliably collected, whereas common RL methods need more. 
For the output action, most RL methods focus on acyclic control, which real-world signal controllers do not support. Most importantly, industry standards require a consistent cycle-flow relationship: non-decreasing and different response strategies for low, medium, and high-level flows, which is ignored by the RL methods.} To narrow the gap between RL methods and industry standards, we innovatively propose to use \ree{industry solutions} to guide the RL agent. Specifically, we design behavior cloning and curriculum learning to guide the agent to mimic and meet industry requirements and, at the same time, leverage the power of exploration and exploitation in RL for better performance. 
\re{We theoretically prove that such guidance can largely decrease the sample complexity to polynomials in the horizon when searching for an optimal policy.} Our rigid experiments show that our method has good cycle-flow relation and superior performance. The code is available \href{https://github.com/AnonymousIDforSubmission/GuidedLight}{\textcolor{blue}{\underline{here}}}.

\end{abstract}

\section{INTRODUCTION}
Traffic signal control (TSC) is a critical chapter to maintain traffic efficiency and safety in smart sustainable mobility \cite{ketter2023information}. 
Recently, employing reinforcement learning (RL) to solve TSC  has become a research hotspot since it does not require too much artificial priori and can improve performance significantly through the paradigm of trial and error. Despite all efforts in this field, there is no real-world deployment of an actual RL-based TSC to date. In fact, it has not yet been proven that RL is even applicable as TSC in a real-world setting. Every investigation so far has been done without considering real-world deployment requirements. In this paper, we aim for a practical RL-based
traffic signal control method, considering the following industry requirements:

\begin{figure}[t]
\centering
\includegraphics[width=0.99\columnwidth]{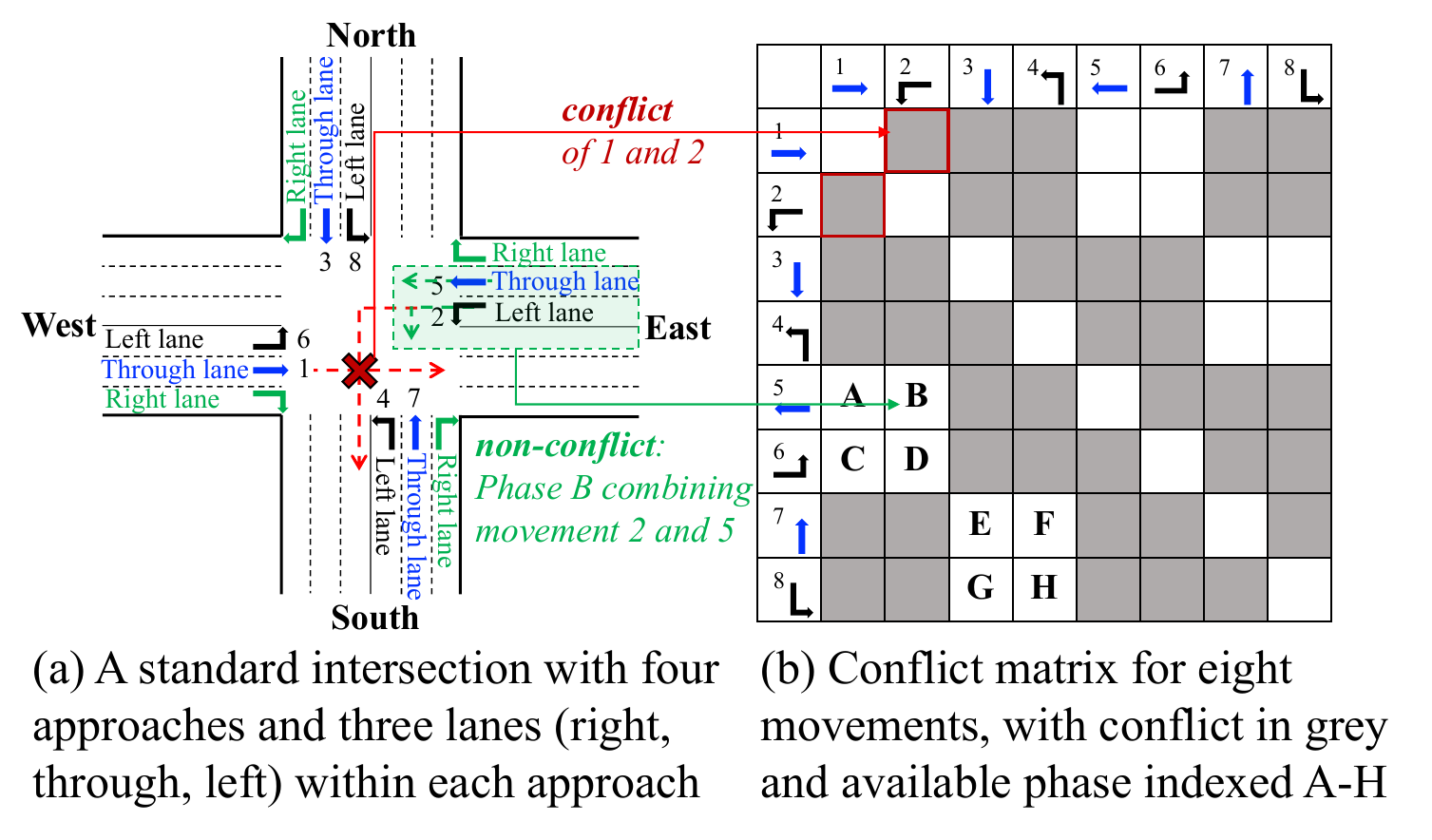}
\caption{Illustration of a standard intersection and phases. \ree{A phase is standardized as a combination of two non-conflicting movements, e.g., phase A is movement 1 and 5 combined. In industry, a four-phase cycle control, i.e., A$\to$D$\to$E$\to$H, is common, representing EW-Though $\to$ EW-Left $\to$ SN-Through $\to$ SN-Left.}}
\label{fig: intersection}
\end{figure}

\begin{figure}[t]
\centering
\includegraphics[width=\columnwidth]{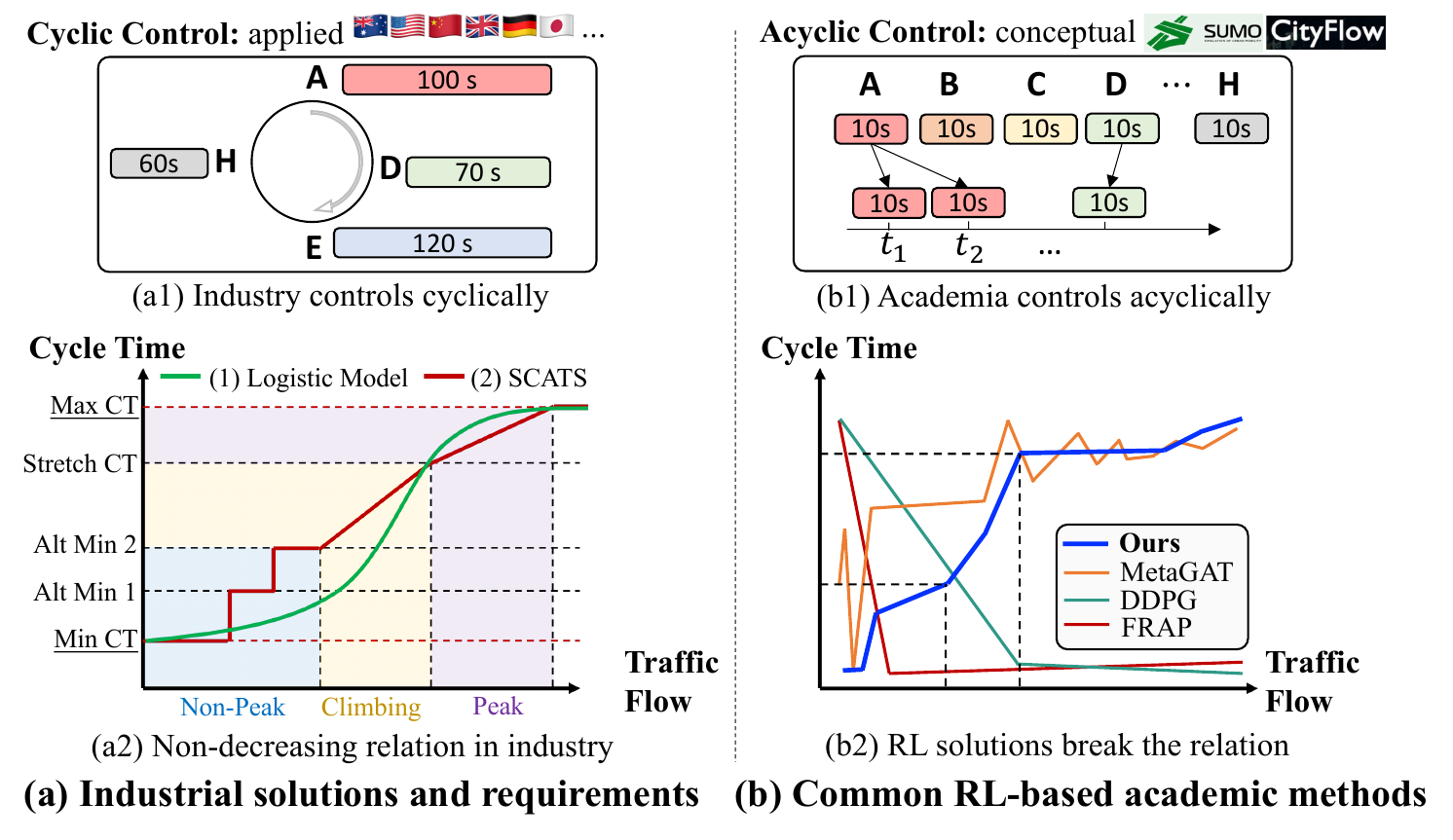}
\caption{The gaps between industry requirements and RL-based academic methods: \ree{\textbf{Industrial requirements:} (a1) they follow cyclic control, e.g., the cycle of phases A-D-E-H (details in Fig. \ref{fig: intersection}), and decide the duration for each phase; (a2) The cycle time should be non-decreasing with traffic flow, encouraging stable drivers' behaviors. It also has three stages: non-peak, climbing, and peak, with different stages featuring various controlling patterns. \textbf{Academic Solutions:} (b1) Common RL-based methods in academia control cyclically, i.e., to choose one phase out of eight phases (A-H) for the next time interval, with each chosen phase running for a fixed interval, e.g., 10 seconds; (b2) The cycle time-traffic flow relation is also diverse and unstable, disrupting drivers negatively. \textbf{Our solution} is the first RL-based cyclic TSC agent, which perfectly follows the non-decreasing relations, with also clear three stages.}}
\label{fig: intro-demo}
\end{figure}

\noindent$\bullet$~\textbf{R1: Input (State)}: \re{The RL-based methods usually need multiple environmental observations, which are rather challenging to obtain in the real world. }
Specifically, they need a combination of queue length, traffic flow, wait or delay time, car position, and so on as the observation space~\cite{van2016coordinated, mousavi2017traffic,wei2018intellilight,zheng2019learning,jiang2022general,lu2023dualight}. However, traffic flow is the only available and stable index in most cities. \re{SCATS, SCOOT~\cite{hunt1982scoot}, and other traditional industrial solutions thus only use traffic flow to provide stable and good enough performance.} 

\noindent$\bullet$~\textbf{R2: Output (Action)}: The learned RL policies do not align with the industry practice. Firstly, Fig. \ref{fig: intersection} shows how traffic signals are usually controlled with the basic concept of ``phase'', i.e., a combination of two non-conflicting traffic movements (more details in Sec \ref{sec: pre}). As shown in Fig.~\ref{fig: intro-demo}(a1), the real-world solution is to directly determine the cycle time for each of the fixed-order four phases, e.g., A-D-E-H, and then repeat, known as ``Cyclic Control''. (Details in Sec \ref{sec: pre}.) In cyclic control, the task is to determine the phase duration for each phase of A, D, E, and H within a cycle. Thus, we can also sum up to get the whole cycle length. Cyclic control is known to be friendly and intuitive for drivers to follow. \re{As depicted in Fig.~\ref{fig: intro-demo}(b1), almost all the RL models \cite{wei2018intellilight,zheng2019learning,jiang2023a,lu2023dualight,lou2022meta,zang2020metalight,el2013multiagent,wong2023deep,2019PressLight,ruan2024coslight,jiang2024x,du2024felight,li2021improved} follow an ``Acyclic Control'', which is still conceptual in simulation systems and not validated in real life: the action is to choose one phase from 8 available phases (with fixed 10 seconds to release) for every 15 seconds. Thus, the resulting phase sequence can be quite random and unstable, causing drivers' confusion and even potential traffic accidents.}

\noindent$\bullet$~\textbf{R3: Cycle-Flow Relation}: \re{Most importantly, the industry requires that for each intersection: 1) the trends of the traffic flow and the cycle time should be positively synchronized, which means the cycle-flow curve shown in Fig. \ref{fig: intro-demo}(a2) should always be non-decreasing, between the required minimal and maximal cycle time. \textcolor{black}{A straightforward explanation is that as the traffic flow regulated by this phase increases, the duration required to fully clear this phase of the traffic flow will also extend. Otherwise, the start time and stop time of the unreleased vehicle will be lost}; 2) moreover, the relation between the traffic flow and the cycle time should be three-stage~\cite{manual2000highway}: as shown in Fig. \ref{fig: intro-demo}(a2), both logistic model \cite{lloyd1967american} and SCATS \cite{lowrie1990scats} show that when the flow is medium (yellow region, climbing up to the peak), cycle time should response very sensitively, and the other regions such as the purple region where the flow is reaching the traffic capacity, the cycle time change should be smooth, to avoid further delay. Mechanisms of R2 and R3 cultivate stable behaviors and response patterns from the driver to the traffic lights and avoid confusion and even potential traffic accidents. In contrast, as shown in Fig. \ref{fig: intro-demo}(b2), the RL-based methods such as DDPG (Deep Deterministic Policy Gradients) \cite{lillicrap2015continuous}, FRAP (short for Flipping and Rotation and considering All Phase configurations) \cite{zheng2019learning}, and MetaGAT \cite{lou2022meta}, have quite unreasonable cycle-flow relations, heavily confusing the drivers.}

In this paper, we present a practical RL model to tackle all three criteria above. We will use the traffic flow as state input and output cyclic phases as action so the trained RL model is campatible with industry practice. Moreover, we also aim to maintain a rigid and rule-based-like cycle-flow relation in Fig. \ref{fig: intro-demo} by combining the traditional solution with the RL method, to obtain a solution that meets industrial needs and has good performance at the same time. \re{In this paper, we propose, GuidedLight, an RL-based agent guided by the ``\ree{industrial solutions}''. Specifically, we use Behaviour Cloning (BC) to encourage the agent to learn from the teacher, e.g., SCATS, and we also let the learning process be gradual and progressive: we adopt curriculum learning \cite{narvekar2020curriculum}, which teaches the agent from easy to advanced}. In summary, the contributions are three-fold:

\begin{itemize}
    \item 
    \ree{\textbf{Our GuideLight lays a significant foundation for the practical implementation of RL-based solutions to be adopted in practice}}. Technically, our RL framework is purely designed in a way that matches the industry standards in terms of state, action, rewards, and cycle-flow relation. 
    \item \ree{To guarantee the rigid cycle-flow relation}, we innovatively propose to use industrial rule-based solutions to guide the agent via behavior cloning and curriculum learning. Moreover, we also theoretically prove that with guidance from the traditional methods, our method guarantees a polynomial sample complexity in the horizon. 
    \item Experiments show that the proposed method not only respects the cycle-flow relations, but also achieves higher performance, thanks to the ``industrial'' guidance and the agent's own exploration and exploitation.
\end{itemize}

The rest of the paper is organized as follows: In Section \ref{sec: lr}, we will review the traditional methods and the RL-based methods for traffic signal control; In Section \ref{sec: pre}, we will give the key background information and preliminaries; Section \ref{sec: method} will officially design the proposed ``GuideLight'' and specifically, Section \ref{sec: theo-proof} will prove that such guidance can largely decrease the sample complexity to polynomials in the horizon when searching for an optimal policy. Section \ref{sec: exp} will present the rigid experiments, Section \ref{sec: dis} discusses generalization, and Section \ref{sec: con} will conclude.

\section{RELATED WORK}
\label{sec: lr}
\subsection{Traditional methods }

Traditional methods are widely used in the real world, most of them \re{are cyclic control, which is friendly and consistent to drivers; however, they} rely on strong assumptions and manually specified rules. Fixed-time control~\cite{roess2004traffic} is one of the earliest cycle-based methods: it configures the traffic signal plan as a fixed cycle length, which is inflexible and cannot automatically adapt its policy to the actual situation. Actuated control model~\cite{fellendorf1994vissim, mirchandani2001real} instead used pre-defined sets of rules to decide whether to adjust the current phase. \re{Later on, methods with analytic solutions were proposed:} Webster~\cite{koonce2008traffic} \re{directly calculates a cycle length \re{using the flow ratio as input} to minimize the delay time: $C= \frac{1.5 \cdot L+5}{1-Y}$
where $L$ is the loss time, and $Y$ is the sum of the critical flow ratio.} Logistic method~\cite{koonce2008traffic} uses the logistic curve to better respond to different types of traffic flow. SCATS~\cite{lowrie1990scats} is the most used one worldwide, which does not use a mathematical model but a set of heuristic rules: it uses a set of simple algebraic expressions to describe the traffic characteristics and operating rules of the current road network. \ree{As shown in Fig. \ref{fig: intro-demo}(a2)'s red curve, SCATS handles (1) non-peak traffic flow with ``stair-like'' rules, with indexes such as \textit{Min CT, Alt Min 1, Alt Min 2}  to be specified, (2) climbing traffic with rapid response up to \textit{Stretch CT}, and (3) peak traffic with a flatter curse, capped by \textit{Max CT}.} All of these traditional methods are simple and interpretable, but their assumption dependencies are relatively straightforward and inflexible. Actual situations may not conform to them, so traditional methods often cannot guarantee optimal results.

\subsection{RL methods.}
In recent years, numerous RL-based TSC methods~\cite{yau2017survey, wei2019survey} have achieved significant outcomes. \re{As mentioned before, most RL-based methods are acyclic control, whose action is to decide which phase to choose.} In~\cite{prashanth2011reinforcement, el2013multiagent, xu2013study, van2016coordinated}, a single agent is employed to manage all intersections within a given scenario. However, due to the joint of their state space or action space, these often encounter issues such as the curse of dimensionality \cite{wong2023deep} and instability \cite{lou2022meta}. A different approach is taken by~\cite{2019PressLight, wei2018intellilight, gao2017adaptive}, which utilizes image-based states for decentralized intersection control. \re{\cite{zheng2019learning}, further proposed FRAP, a widely followed method for acyclic control. It inputs the traffic features of all eight movements, combines two non-conflicting movements into a phase (8 phases obtained in total), and uses deep Q-learning to decide which phase will be selected for the next time step.} Building upon this,~\cite{jiang2023a} further presents a General Plug-In (GPI) module, \re{where any arbitrary and various intersections can be mapped into the standard one, thus having a unified state and action space for various intersections.} This facilitates the co-training of large-scale scenarios.

\re{Nonetheless, this acyclic control has quite a wide gap to real-world applications, and a major reason is the hardware: most traffic signal controllers only support cyclic phase plans.}
A few methods have been tried to solve it. For instance, \cite{casas2017deep} used DDPG \cite{lillicrap2015continuous} to determine each phase duration in the next cycle. Nonetheless, this method requires the total cycle time to be fixed. Alternatively, \cite{xu2021hierarchically} uses a strategy of deciding whether to remain in the present phase or transition to the subsequent phase during each short interval. However, this approach neglects various practical constraints, such as ensuring that the discrepancy between the durations of two consecutive cycles is not excessively large. An innovative strategy is presented in~\cite{liang2019deep}, which modifies one phase's duration during each timestep. This approach accommodates certain constraints and allows for dynamic adjustments in both the complete cycle duration and the duration of individual phases. However, when confronted with a limited variety of observations, e.g., only flow is available and a preferred rigid cycle-flow relationship, this method may be incompetent.

\section{PRELIMINARIES}
\label{sec: pre}
Some necessary domain information is given here. 

\textbf{Definition 1. Intersection} is where two or more roads cross. A 4-arm intersection is shown in Fig. \ref{fig: intersection}, where each arm (N, S, W, E) has entering lanes and exiting lanes.  

\textbf{Definition 2. Traffic Movement} is defined as the direction in which a vehicle crosses from an entering lane to an exiting lane, including go-through, turn-right, and turn-left in each arm. In Fig~\ref{fig: intersection}, the numbers 1-8 represent the eight movements controlled by the signal (right movements are usually free).

\textbf{Definition 3. Phase} is a combination of two non-conflicting movements that could be released together. For instance, movements 1 and 2 conflict, thus not being able to form a phase; but movements 1 and 5 can form \textit{W-E-through} phase, denoted as phase-A. \ree{In reality, not all the intersections have all four legs with all movements and phases; in that case, zero-padding will be used to mask the missing movement or phase.}

\textbf{Definition 4. Cyclic Control}
In real-world solutions like SCATS, four phases are usually controlled in a cyclic way: $A \to D \to E \to H$ and then repeat, where in each phase, the phase duration needs to be decided. The cycle time is the summation of all phase durations plus red and yellow time. Our RL method follows the four-phase cyclic control.

\begin{figure*}[h!]
\centering
\includegraphics[width=0.99\textwidth]{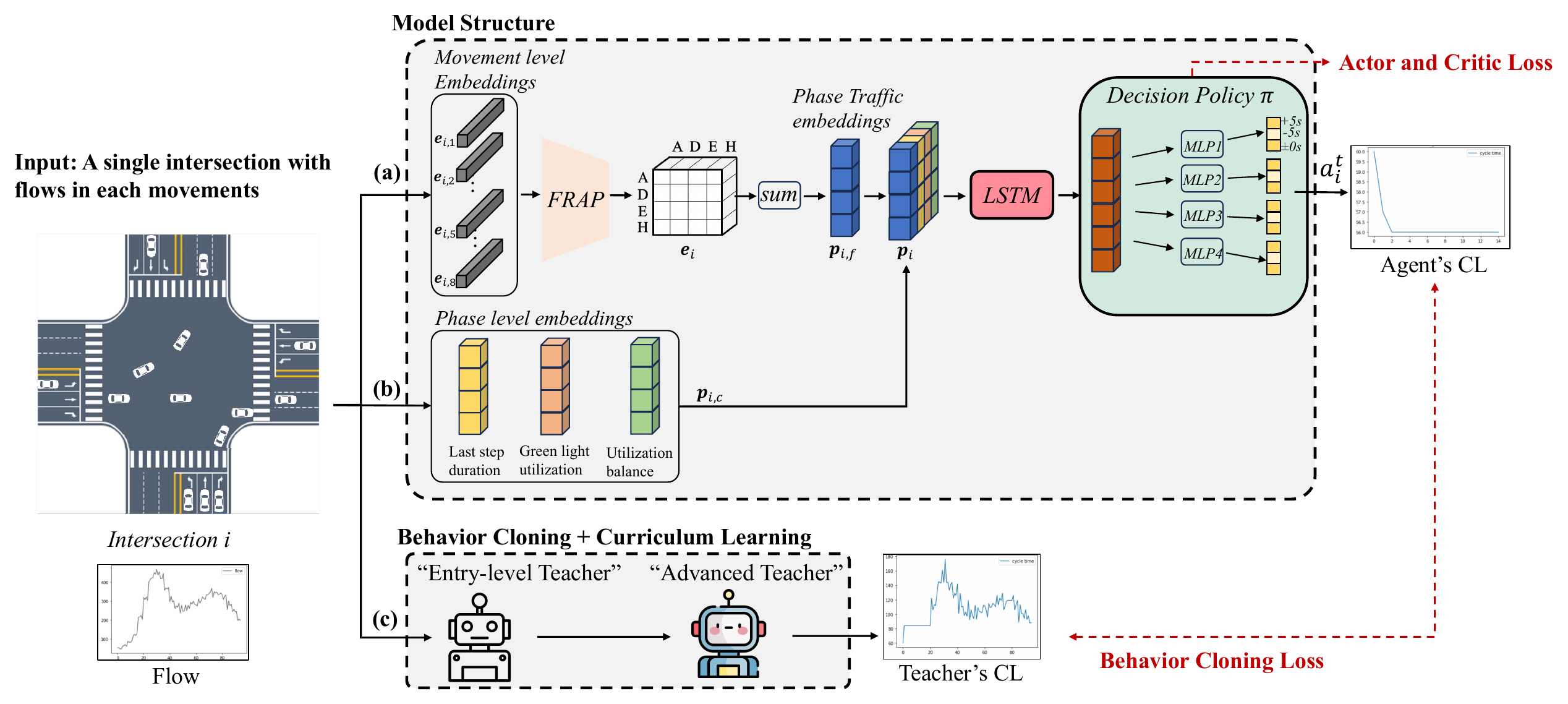}
\caption{Illustration of our proposed GuidedLight. \ree{One agent controls one intersection. (a) For each intersection, it will observe movement-level features related to traffic dynamics and then use FRAP's aggregating module to aggregate two non-conflicting movements’ embeddings into one phase of traffic embedding. (b) Combining phase traffic embedding with other phase-level features, we get a wholesome embedding for a phase and then input them into LSTM and Actor-Critic. (c) Teachers such as linear controller and industrial controller SCATS are adopted, via Behavior Cloning and Curriculum Learning, to guide the agent's cycle length (CL) in mimicking the teacher's CL.}}
\label{fig: network}
\end{figure*}

\section{METHOD}
\label{sec: method}
Sec. \ref{section: RL Setup} will introduce the state space,  action space, and reward design. Sec. \ref{section: Network} introduces the network design. Sec. \ref{section: Training} introduces the training based on behavior cloning and curriculum learning. Sec. \ref{sec: theo-proof} will theoretically prove the improved sample complexity with horizon $H$.

\subsection{RL Setup} \label{section: RL Setup}

Generally, the single-agent TSC problem is formulated as a Markov decision process $\langle \mathcal{S},\mathcal{A}, r,\gamma, \pi \rangle$. \textcolor{black}{Every agent controls an intersection independently.} 

\re{To perfectly match our proposed algorithm with real-world situations, the state, action, and reward are all carefully designed.

\textbf{State $\mathcal{S}$}:} Only traffic flow and last step action-based observations (e.g., 
sampled every 5min or 15min) are chosen since commonly they are the only available index; \ree{Thus, the state space is continuous.} 

\textbf{Action $\mathcal{A}$}: Instead of choosing the next phase in the acyclic control, our action is designed to directly output the phase duration for each phase in a cyclic control. Besides, there are two more constraints: the total cycle time should change smoothly and should stay within the minimal and maximal thresholds. Thus, we translate the action into 3 categories, \ree{thus being discrete actions}, which are adding/cutting 5 seconds (+5s/-5s) or non-changing (0s) for each phase: such a design frees us from the computation burden of the original purely continuous action space, and the small change of 5s on each phase ensures the smoothness of cycle time. To consider the constraint of the total cycle time staying within the minimal and maximal thresholds, once the cycle time hits the min (max) threshold, the action of -5s (+5s) will be masked. Thus, we have:

\begin{equation}
    a =
\begin{cases}
    \{ + 5s, - 5s, 0s\},& \text{if CT} \in [\text{min-threds}, \text{max-threds}] \\
    \text{masked},              & \text{otherwise}
\end{cases}
\end{equation}

\textbf{Reward $r(s,a)$}: we define the reward based on throughput $v$ ($\uparrow$, the higher the better), queue length $l$ ($\downarrow$), and two factors that matter in the industry, i.e., green-light utilization rate $gr$ ($\uparrow$), and green imbalance $gi$ ($\downarrow$). Our reward function is the weighted sum of the four factors, i.e., 

\begin{equation}
    r=w_{v} \cdot v + w_{l} \cdot l + w_{gr} \cdot gr + w_{gi} \cdot gi 
\end{equation}

\textcolor{black}{with weights tuned and specified  in Sec.~\ref{sub:Implementation} 
}. 
Throughput $v$ is the number of vehicles passing through the intersection per unit of time. Queue length measures the length of the vehicle queue at the intersection. Greenlight utilization rate measures the portion of greenlight that is really used to release vehicles, which is inferred as follows for each phase:

\begin{equation}
gr = \frac{v \times 2.5s}{\text{phase duration}}
\end{equation}

where $v$ is the through-put of this phase and 2.5 seconds (a common setting in traffic control manual \cite{koonce2008traffic}) is usually the time for a vehicle to pass the intersection. Green imbalance $gi$ evaluates the standard deviation of all the phases' $gr$:
\begin{equation}
gi = \text{standard deviation}(\{gr\}_{\text{all phases}})
\end{equation}

It is worth mentioning that in the reward, we use metrics 
beside traffic flow, such as queue length, the reason is: during training in the simulation, the queue length is obtainable and it is beneficial to be part of the reward to give the agent better-informed feedback, so that when being deployed in the real world, even only traffic flow is observable, but agent's action can still tend not to cause long-queue.

\textbf{Optimal policy} $\pi^*(a|s)$: At each intersection $i$ and each timestep $t$ the goal of the agent is to find a policy $\pi(a|s)$ that maximizes the expected return $G_t := \sum_t \mathbb{E} [\gamma^tr_t^i]$, where $\gamma$ is the discount factor.
\begin{equation}
\pi^*(a|s) = \arg \max_{\pi} \sum_t \mathbb{E} [\gamma^tr_t^i]
\end{equation}


\subsection{Network design}\label{section: Network}

The model is shown in Fig. \ref{fig: network}. When deciding for $t+1$, we utilize the movement-level and the phase-level features from $t$. Then the fused feature is fed into LSTM~\cite{hochreiter1997long} to capture long-term decision dynamics, and finally, the actor-critic network makes the action~\cite{jiang2023a}. 

\textbf{For the movement-level feature}: 
the $i$-th intersection in the scenario, the $m$-th movement ($m\in \{1,\ldots,8\}$) observes $K$ features. In our case, $K=3$ includes only traffic dynamics, i.e., traffic flow, and two additional static indices, i.e., road capacity and indicator of whether the movement exists.
For any traffic movement,  we get its embedding, where $||, \textit{Sigmoid}(\cdot), \textit{MLP}(\cdot)$ are concatenation, activation function, and multilayer perceptron, respectively. 
\begin{equation} \label{eq: movements}
    \bm{e}_{i,j} = ||_{k=1}^K\textit{Sigmoid}(\textit{MLP}_k(s_{i,m,k})),
\end{equation}

Then, we use part of FRAP to first aggregate two non-conflicting movements' embeddings into a phase embedding, and then also to embed the phase conflict information into the phase embeddings. \textcolor{black}{(1) Add: FRAP adds the embedding of movements as a phase representation. For the $p$-th phase which consists of movements $j, j'$, $p \in [A, B, \dots,H]$: $\bar{\textbf{e}}_{i, p} = \textbf{e}_{i, j} + \textbf{e}_{i, j'}$, where the two movements $j, j' \in [1, 2, \dots, 8]$.
(2) Phase-pair representation: Once obtaining the phase embedding, a phase-pair representation is constructed to capture the pairwise relations for competition: $\hat{\mathbf{e}}_i = \bar{\textbf{e}}_{i, l} \| \bar{\textbf{e}}_{i, l'}$.
(3) Phase competition: To avoid phase conflict, pairwise competition scores \cite{zheng2019learning} are learned as a competition mask, denoted as $\boldsymbol{\Omega}$. Thus, the phase-pair representations $\hat{\mathbf{e}}_i$ will go through $1 \times 1$ convolution and multiply with the phase competition mask to yield the masked phase-pair embedding: $ \bm{e}_{i} = \textit{Conv}_{1 \times 1} ( \hat{\mathbf{e}}_i ) \otimes \boldsymbol{\Omega}$, and $\otimes$ is Kronecker product. We denote the whole FRAP module as $\textit{FRAP}(\cdot)$} \cite{zheng2019learning}:
\begin{equation}
    \bm{e}_{i} = \textit{FRAP}(\bm{e}_{i,1},\ldots,\bm{e}_{i,8}), \text{where }  \bm{e}_{i} \in \mathbb{R}^{4 \times 4 \times d_{f}},
\end{equation}
$\bm{e}_{i}$ is 4-phase embedding tensor (phase A-D-E-H), with embedding dimension as $d_{f}$. 
We sum it along the row direction and get the phase embedding with traffic \underline{f}low information.
\begin{equation}
    \bm{p}_{i,f} = \textit{sum}(\bm{e}_{i}), \text{where } \bm{p}_{i,f}\in \mathbb{R}^{4 \times d_{f}}
\end{equation}

\textbf{For the phase-level feature}: $\hat{s}$ denotes the phase-level raw observations with $\hat{K}$ features in these observations. In our case, $\hat{K}=3$, including more \underline{c}ontexts of each phase's duration at the last cycle time, greenlight utilization rate, and green imbalance. We use each MLP to embed them.
\begin{equation}
    \bm{p}_{i,c} = ||_{\hat{k}=1}^{\hat{K}} \text{MLP}(\hat{s}_{\hat{k}}),
\end{equation}

We concatenate $\bm{p}_{i,f}$ and $\bm{p}_{i,c}$ as the intersection $i$'s feature:
\begin{equation}
    \bm{p}_{i} = ||(\bm{p}_{i,f}, \bm{p}_{i,c})
\end{equation}
Lastly, we adopt the LSTM to perceive long-term historical states, so that the agent can use historical trends to achieve better consistency between phase duration and traffic flow. \textcolor{black}{We take the phase-level feature $\bm{p}_{i}^t$ at this time step $t$ of the intersection $i$ and last time step LSTM hidden state $c^{t-1}$ to the LSTM and get the $h_i^t$ and $c^{t}$.}
\begin{equation}
    h_i^t, c^{t} = \textit{LSTM}(\bm{p}_{i}, c^{t-1}),
\end{equation}
For the decision policy $\pi$, it receives the $h_i^t$ as input and outputs the action $a_i^t$.
\begin{equation}
    a_i^t = \pi(\cdot|h_i^t).
\end{equation}
\re{As mentioned in Sec \ref{section: RL Setup} about action, $a_i^t$ is defined as +5s/-5s/0s} for each phase. 

\subsection{Training}\label{section: Training} 

Our experiments show that solely optimizing the agent using the RL rewarding mechanism only yielded unsatisfactory results regarding the industry requirement (consistent cycle-flow relation). 
The reasons are two-fold: (1) the agent only perceives the traffic flow from the dynamic road network. This restricted input hampers the agent's ability to explore and identify a superior policy. (2) moreover, the reward is hard to align with the stringent requirements of the industry. 

\textbf{Behavior Cloning}: To address this predicament, we employ BC to steer the agent's actions: \re{namely, we employ the traditional and market-proved solutions to guide our agent.} This approach substantially narrows the agent's exploration domain, facilitating faster optimization \cite{chang2023learning}.

\re{However, the well-developed solutions such as SCATS directly output a continuous phase duration, and ours is a discrete action of +5s/-5s/0s. How can we design a loss to penalize the difference between the two? We translate them into logits.} The action of the $p$-th phase from our agent's decision policy $\pi$ can be sampled from below logits:
\begin{equation}
    \bm{l}_{i,p}^t = \text{MLP}_p(h_i^t), \text{where } \bm{l}_{i,p}^t \in \mathbb{R}^{3},
\end{equation}
The label of BC guidance can be generated as:
\begin{equation} \label{eq: expert}
\small
    \hat{\bm{l}_{i,p}^t} = \left\{ 
    \begin{array}{lc}
        0 & \bm{E}_{i,p}^t - 5 \geq T_{i,p}^{t-1}\\
        1 & \bm{E}_{i,p}^t + 5 \leq T_{i,p}^{t-1}\\
        2 & \text{otherwise} \\
    \end{array}
\right.
\end{equation}
\noindent where $\bm{E}_{i,p}^t$ is the phase duration decided by the expert model of at time $t$, the $T_{i,p}^{t-1}$ is the duration time at last time step, the 0/1/2 represents the ground truth is  +5s/-5s/0s respectively. Lastly, we employ the \textit{Cross Entropy} as the BC loss:
\begin{equation}
    \mathcal{L}_{BC} = \textit{CrossEntropy}(\bm{l}_{i,p}^t, \hat{\bm{l}_{i,p}^t}), 
\end{equation}


\textbf{Curriculum Learning}: Delving deeper into the BC process, we adopt the Curriculum Learning approach to ensure the agent's progressive learning trajectory. 
\re{We gradually introduce teachers from easy to advanced to generate the label in Eq. (\ref{eq: expert}).}
\begin{itemize}
    \item \textbf{Easy guide}: we choose a linear model that posits a straightforward linear correlation between cycle duration and traffic flow. \re{$\text{Phase duration} \approx 0.35 \times v$, where $0.35$ is roughly estimated from regression based on large amounts of offline data.}
    \item \textbf{Medium guide}: logistic curve based on \cite{koonce2008traffic} could be used, where the cycle time and traffic flow follow a logistic relation between the \textit{Max CT} and \textit{Min CT}, as shown in Fig. \ref{fig: intro-demo}(a2).
    \item \textbf{Advanced guide}: In contrast, SCATS exhibits a more intricate behavior, with the three-stage cycle-flow relations, as shown in Fig. \ref{fig: intro-demo}(a2), thus being the most advanced guide.
\end{itemize}


\ree{\textbf{Total Loss:} The Behavior Cloning and Curriculum Learning are combined with the Actor-Critic to formulate} the total training loss:
\begin{equation}
    \mathcal{L} = \alpha\mathcal{L}_{Actor} + \beta\mathcal{L}_{Critic} + \kappa \mathcal{L}_{BC}. 
\end{equation}
where $\alpha, \beta, \kappa$ are tuning parameters. The Actor loss and the Critic loss are the same as Proximal Policy Optimization (PPO)~\cite{schulman2017proximal}.

\begin{algorithm}[t]
\caption{GuideLight training process}
\label{alg:algo}
\textbf{Input}: A set of target intersections $\mathcal{I}$; training episodes \textit{E}, a set of teacher algorithms \textit{B}; 


\textbf{Output}: Optimized policy $\pi$;
\begin{algorithmic}[1]
\FOR {episode=1, $\ldots, \textit{E}$}
\STATE Initialize LSTM hidden state $c^0$ for each intersection and clean buffer $\mathcal{D}\leftarrow \emptyset$.

\FOR {each time step $t$}
\STATE Record each raw observation of each intersection to $\mathcal{D}$;
\STATE Get the current time step intersection feature according to Eq. (10) and use LSTM to perceive long-term historical information according to Eq. (11);
\STATE Get and take action $a_i^t$ according to Eq. (12);
\STATE Get the rewards $\mathbf{r}_t^i$ and the next time step raw observation;
\ENDFOR
\FOR{each step in training steps}
\STATE sample minibatch data from $\mathcal{D}$;
\STATE Select one teacher from $B$ according to the training process and introduce the teacher label according to Eq. (14) ;
\STATE Computer $\mathcal{L}$ according Eq. (16) and update Network parameter;
\ENDFOR
\ENDFOR
\end{algorithmic}
\end{algorithm}

\subsection{Theoretical Analysis}
\label{sec: theo-proof}

\ree{Will introducing such guidance even speed up learning?} We prove that our approach, under the assumption of guided policy efficacy, can significantly reduce the sample complexity from exponential in the horizon to polynomial.

\textbf{Theorem 1.} \textit{For 0-initialized $\epsilon$-greedy, where initial estimates for action probabilities start at zero,  there exists an MDP instance where the sample complexity scales exponentially with the total time horizon H to identify a policy with a suboptimality less than 0.5. \cite{koenig1993complexity}}

\textbf{Assumption 1.} 
\textit{Assume that the optimal policy as $\pi^\star$, with its visited state distribution as ${ d  ^ {  \star  }} $.  
The policy updated based on the guidance loss is denoted as $\hat{\pi^g}$ with its visited state distribution as ${ d  ^ {  \hat{g}  }} $. 
We make the assumption that $\hat{\pi^g}$ cover the states visited by $\pi^\star$ subject to an upper limit:
$$\operatorname { sup } _ { s } \frac { d  ^ {  \star  } (  s  ) } {  d  ^ {  \hat{g}  }   (  s  ) } \leq C$$}


This ratio, also referred to as the distribution mismatch coefficient, is used in gradient descent methods~\cite{agarwal2021theory}. 

We explain the advantage of guidance based on an online contextual bandit scenario \cite{tekin2015distributed}, and we look at \textit{regret}, a measure of the difference between the reward that could have been obtained by always taking the best possible action and the reward that was actually obtained by the policy used.

\textbf{Assumption 2.} \textit{We assume that there is a guarantee for the exploratory policy. 
Within the confines of an online contextual bandit scenario characterized by a stochastic context $s \sim p_0$ and a reward function $r(s, a)$ spanning $[0, R]$, an exploratory policy exists in every round, such that the total regret is bounded:
$$\sum _ { t = 1 } ^ { T } \mathbb{E} _ { s \sim p _ { 0 } } [ r ( s , \pi ^ { \star } ( s ) ) - r ( s , \pi ^ { t } ( s ) ) ] \leq f ( T , R )$$}


This exploratory policy guarantee is a prevailing assumption in extensive literature, whether in tabular methods~\cite{langford2007epoch} or in methods similar to ours using general function approximation~\cite{simchi2022bypassing, krishnamurthy2020contextual}.


Based on the Performance Difference Lemma proposed by Kakade et al. \cite{kakade2002approximately}, we can establish a relationship between the iterations $I$ and the discrepancy between the optimal value function $V _ { 0 } ^ { \star }$ and the current value function $V _ { 0 } ^ { \pi }$.
\begin{align}
& E _ { s _ { 0 } \sim d _ { 0 } } [ V _ { 0 } ^ { \star } ( s _ { 0 } ) - V _ { 0 } ^ { \pi } ( s _ { 0 } ) ]  \nonumber \\
& = \sum _ { i = 0 } ^ { I } E _ { s \sim d _ { i } ^ { \star } } [ Q _ { i } ^ { \pi } ( s , \pi _ { i } ^ { \star } ( s ) ) - Q _ { i } ^ { \pi } ( s , \pi _ { i } ( s ) ) ]
\end{align}

Consequently, we can derive an upper bound for regret. Let $T$ denote the total exploration steps and $E$ represent the number of epochs after which the policy is updated. Hence, the final number of policy iterations is $T/(EH)$.
\begin{align}
& \mathbb{E} _ { s _ { 0 } \sim d ^ { \star } } [ V _ { 0 } ^ { \star } ( s _ { 0 } ) - V _ { 0 } ^ { \pi } ( s _ { 0 } ) ] \\
& = \sum _ {i= 0 } ^ { T/EH } \mathbb{E} _ { s \sim d _ {i} ^ { \star } } [ Q _ {i}  ( s , \pi _ {i} ^ { \star } ( s ) ) - Q _ {i}  ( s , \pi _ {i} ( s ) ) ]    \\
&  { \leq } C \cdot \sum _ {i= 0 } ^ { T/EH } \mathbb{E} _ { s \sim d _ {i} ^  { g } } [ Q _ {i}  (   s  ,  { \pi } _ {i} ^ { \star } ( ( s ) ) ) - Q _ {i}  (   s  , \pi _ {i}  ( s ) ) ] \label{eq:ineq1}\\ 
&  { \leq } C \sum _ {i= 0 } ^ { T/EH } f ( H, R )  \label{eq:ineq2}
\end{align}
In the derivation, inequality \eqref{eq:ineq1} arises from Assumption 1, while inequality \eqref{eq:ineq2} is predicated on Assumption 2. 
Drawing on the results presented~\cite{uchendu2023jump}, it can be inferred that under the setting of asymptotic functions, $f ( H, R ) = R \cdot ( A \mathcal{E} _ { F } ( H ) ) ^ { 1 / 2 }$, where $\mathcal{E}$ is a polynomial function.
Consequently, the rate of our algorithm is up to a factor of $C \cdot \textit{poly}(H)$ \cite{simchi2022bypassing}.


\section{EXPERIMENT}
\label{sec: exp}
\subsection{Towards Real-world Dataset and Simulation}
\begin{figure}[t]
\centering
\includegraphics[width=\columnwidth]{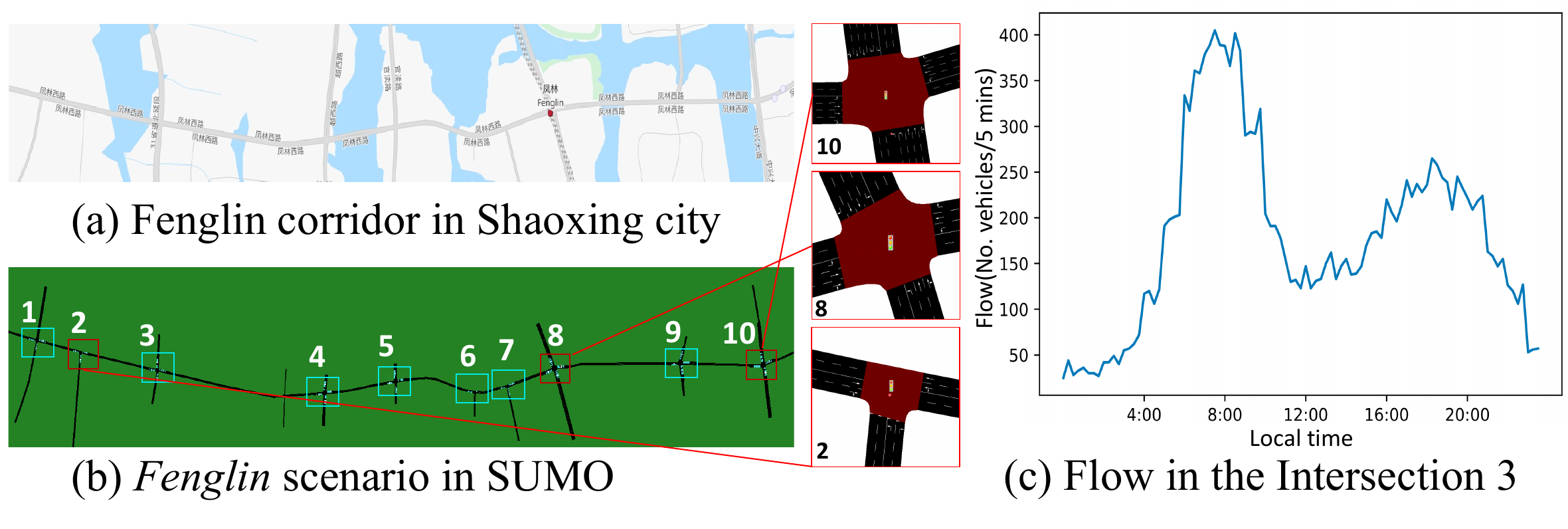}
\caption{Illustration of Fenglin scenario.}
\label{fig: dataset}
\end{figure}

\re{Several efforts have been made to ensure that our evaluation scenario is as consistent as possible with the real-world scenario. (1) \textbf{Road Network}: we built one simulation scenario in SUMO based on a real-world case: Fenglin West Road, Shaoxing, Zhejiang, as shown in Fig. \ref{fig: dataset}, which is a corridor with 10 intersections to be signal-controlled. 
More details are in \cite{jiang2023a}; (2) \textbf{Traffic flow}: most SUMO scenarios only support 1-hour-long flow data, which is not enough to train agents to learn to respond to all levels of flow, i.e., the 3-stage cycle-flow relations shown in Fig. \ref{fig: intro-demo}. Thus, we collect the real 24-h flow data, e.g., shown in Fig. \ref{fig: dataset}(c), and based on the real-world traffic flow distribution, we generated 1400 different flow patterns for each intersection, maximally enhancing the agents' generalizability; (3) \textbf{More configurations}: more detailed designs such as aligning the SUMO's flow sampling rate with the real-world sensors (e.g., every 5min or 15min), following the pre-defined 4-phase plan (i.e., Phase A-D-E-H), supporting the output of phase duration from agent (instead of choosing the next phase), and so on, have been made. The dataset is open-sourced \href{https://github.com/AnonymousIDforSubmission/GuidedLight}{\textcolor{blue}{\underline{here}}} for the community.}

\subsection{Benchmark Methods}

We compare our GuideLight with both traditional and RL-based methods:

\begin{table*}[t]
\begin{center}
\caption{Main Results. The best \textbf{boldfaced} and second best \underline{underlined}. Our method always achieves the top 2.}
\resizebox{0.95\textwidth}{!}{
\begin{tabular}{ccccccc}
\hline
 &  FTC  & SCATS & DDPG &  FRAP$^*$ & MetaGAT$^*$ & GuideLight (Ours) \\
\hline 
Throughput   & $155.95_{\pm 0.08}$    & $\textbf{161.86}_{\pm 0.31}$ &  $158.34_{\pm 0.17}$   & $113.15_{\pm 0.03}$ & $141.24_{\pm 0.05}$  & \underline{159.63} $_{\pm 0.28}$ \\
\hline
Green Utilisation     & $87.57_{\pm 0.01}$   & $92.19_{\pm 0.02}$ & \underline{117.74}$_{\pm 0.01}$ & $103.02_{\pm 0.02}$ & $98.32_{\pm 0.01}$ & $\textbf{121.80}_{\pm 0.02}$ \\
\hline
Green Imbalance   & $\textbf{-52.60}_{\pm 0.01}$  & $-57.04_{\pm 0.01}$ & $-58.57_{\pm 0.01}$ & $-71.57_{\pm 0.01}$ & $-60.67_{\pm 0.01}$ & \underline{-56.03}$_{\pm 0.01}$ \\
\hline
Queue Length    & $-82.82_{\pm 0.02}$ & $-61.30_{\pm 0.22}$ & $-71.05_{\pm 0.12}$  & $-265.32_{\pm 0.13}$ & \underline{-58.91}$_{\pm 0.12}$ & $\textbf{-55.66}_{\pm 0.29}$ \\
\hline
All             & $108.09_{\pm 0.09}$ & $135.71_{\pm 0.33}$ & \underline{149.47}$_{\pm 0.20}$ & $-120.33_{\pm 0.15}$ & $119.98_{\pm 0.15}$ &  $\textbf{169.74}_{\pm 0.39}$ \\
\hline
\end{tabular}
}
\label{tab2}
\end{center}
\end{table*}

\noindent \textbf{Traditional Methods:}
\begin{itemize}
    \item \textbf{Fixed Time Control} (FTC)~\cite{roess2004traffic} with random offset executes each phase within a loop, utilizing a pre-defined phase duration.
    \item \textbf{SCATS}~\cite{lowrie1990scats} The most widely used conventional traffic control algorithms based on real-time traffic data, road characteristics, and manually designed operational rules.
\end{itemize}

\noindent \textbf{RL-based Methods:}
\begin{itemize}
    \item \textbf{DDPG} \cite{casas2017deep}  A robust RL model widely employed across diverse problem domains and settings, which is meticulously tailored to suit our setting.
    \item \textbf{FRAP$^*$} \cite{zheng2019learning}: It is the ground-breaking acyclic RL based on phase competition. We implement it in a cyclic way based on ours without BC.
    \item \textbf{MetaGAT$^*$} \cite{lou2022meta} combines contextual meta-reinforcement learning based on GRU to improve generalization and GAT for cooperation. We also implement it in a cyclic way. 
    \item \textbf{GuidedLight} Our proposed method in this paper. We used the linear guide and SCATS guide.
\end{itemize}

\subsection{\textcolor{black}{Implementation details of GuidedLight}}\label{sub:Implementation}

The implementation details of GuidedLight are summarized in Table \ref{table:Implementation details of GuidedLight}. The parameters are fine-tuned mainly based on grid search.

\begin{table}[t]
\small
\centering
\caption{Implementation details of GuidedLight}
\resizebox{0.99\columnwidth}{!}{
\begin{tabular}{ll}
\hline
Items                                   & Details \\ \hline
Learning rate                           & 5e-4    \\
Actor loss coefficient $\alpha$                  & 1       \\
Critic loss coefficient $\beta$                 & 1       \\
Behavior cloning loss coefficient $\kappa$      & 0.5    \\ \hline
Weight of Green-light utilization rate in Reward $w_{gr}$                            & 1   \\
Weight of Green imbalance in Reward $w_{gi}$                                & -1   \\
Weight of Throughput in Reward $w_{v}$                            & 4e-2   \\
Weight of Queue length in Reward $w_{l}$                                & -1e-3   \\ \hline
\end{tabular}
}
\label{table:Implementation details of GuidedLight}
\end{table}

\subsection{Evaluation metrics}
We use two metrics to evaluate these methods.  
\re{Firstly, we evaluate the results based on traffic flow, queue length, and two factors that matter in the industry, i.e., \textit{green-light utilization rate}, and \textit{utilization balance}}. These factors form our reward function, which our agent receives as input. Here, traffic flow refers to the volume of vehicles passing through the intersection per unit of time. Green-light utilization measures the proportion of effective time when vehicles pass through the intersection during the green light phase. Utilization balance evaluates the level of balance among the four phases of the intersection. Queue length measures the length of the vehicle queue at the intersection.

Additionally, we visualize the synchronization between cycle time and traffic flow. In practice, this is an important indicator of the TSC system. Cycle time tends to increase within a bounded range as traffic flow increases.

\subsection{Results}

\subsubsection{Main Results}

\begin{figure*}[t]
\centering
\includegraphics[width=\textwidth]{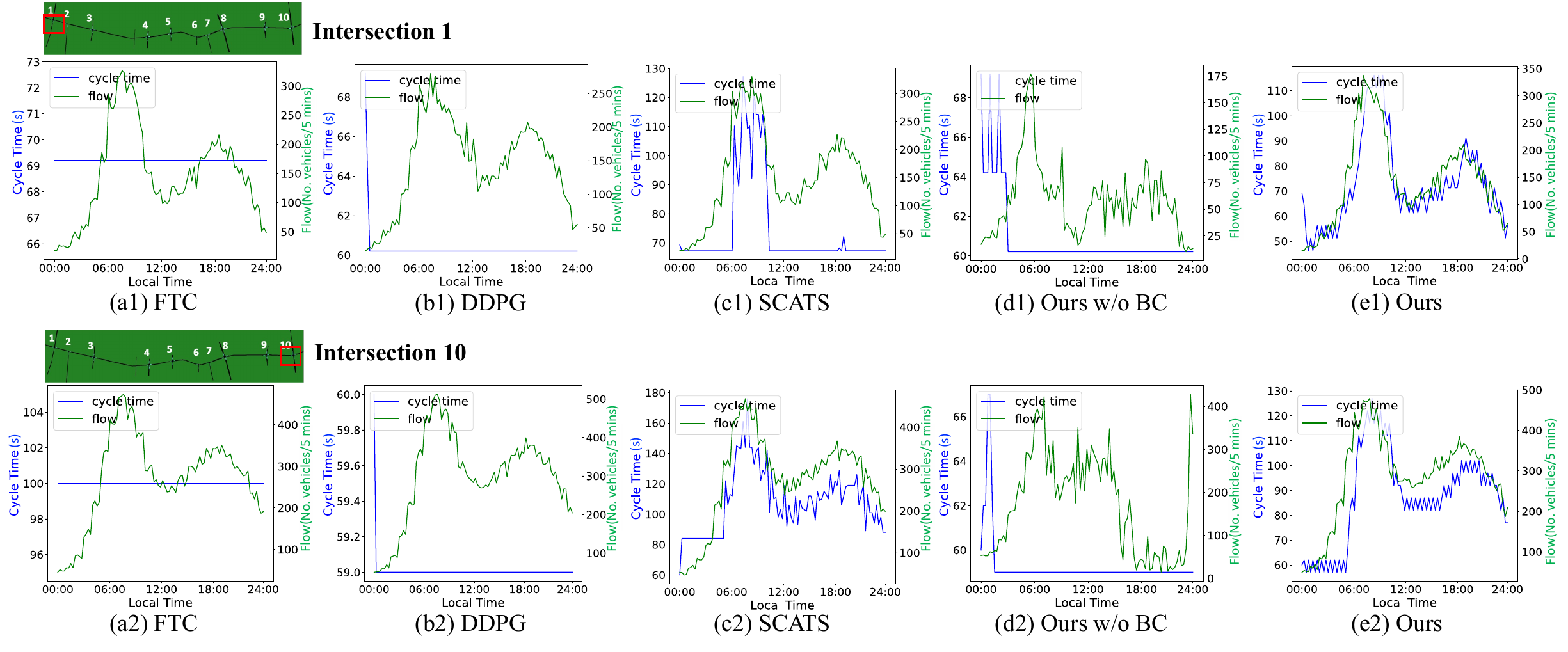}
\caption{The synchronization between \textcolor{blue}{cycle time} and \textcolor{Green}{traffic flow} in \textbf{Fenglin dataset}. As in (e2), \ree{GuidedLight outputs the control policies with cycle time perfectly following the flow and thus meets industry requirements. Fig \ref{fig: intro-demo}(b2) summarizes the overall cycle time-traffic flow relation.}}
\label{fig:syn}
\end{figure*}

Quantitative metrics are displayed in Table \ref{tab2}, demonstrating that our method consistently attains the highest overall scores, showing substantial advantages over other algorithms. 
Particularly, in the crucial metric of green-light utilization rate, which is highly important in real-world applications, our algorithm markedly surpasses both the SCATS and FTC methods.
Although our approach only achieves the second-best throughput and green imbalance, the gap between our performance and the best results in these categories is relatively minor. Detailed analysis is as follows:

\ree{
\textbf{Behavior Cloning and Curriculum Learning helps GuidedLight learn.} Via the delicately designed heuristics, such as the three stages and parameter specifications, SCATS achieves the highest throughput, being the most efficient controller. Thanks to our Behavior Cloning and Curriculum Learning, GuidedLight successfully mimics SCATS's behavior and achieves the second. As illustrated in Fig. \ref{fig: intro-demo}(b2), our GuidedLight also achieves a non-decreasing cycle time-traffic flow relation, and good-shaped three-stage controlling behavior for non-peak, climbing, and peak traffic.

\textbf{RL-based agent boosts GuidedLight's performance in other metrics.} Through the Actor-Critic-based agent, GuidedLight further optimizes industrial-concerned metrics such as green utilization rate and green imbalance, as well as the queue length, crowning our GuidedLight as the best overall performer.

\textbf{Academic solutions based on RL have rather discounted performance and confusing cycle time-traffic flow relations.} The well-recognized top performers, FRAP and MetaGAT, when applied in practical industrial settings, suffered performance drops and cannot beat industrial solutions such as SCATS and ours. This is because their states, actions, and rewards are all derailed from being practical. Furthermore, as illustrated in Fig. \ref{fig: intro-demo}(b2), they even have decreasing cycle time-traffic flow relation, meaning higher traffic demand can trigger shorter cycle time and vice versa, which can be rather confusing for drivers. 
}

\subsubsection{Visualization of Synchronization} 

We visualize the synchronization of cycle time and traffic flow in Fig. \ref{fig:syn} based on two intersections. It shows that our method and the SCATS system achieve notable synchronization; namely, the cycle time co-varies responsively with fluctuations in traffic flow. On the contrary, the FTC and DDPG approaches remain indifferent to such dynamics. 
Synthesizing the aforementioned results, it becomes evident that RL algorithms tend to outshine traditional methods in terms of quantitative evaluation scores. However, the conventional SCATS method demonstrates superior synchronicity of cycle time and flow. Our method, underpinned by BC guidance, excels in quantitative scoring and synchronicity, thus delivering an optimized performance.

\begin{figure}[t]
\centering
\includegraphics[width=\columnwidth]{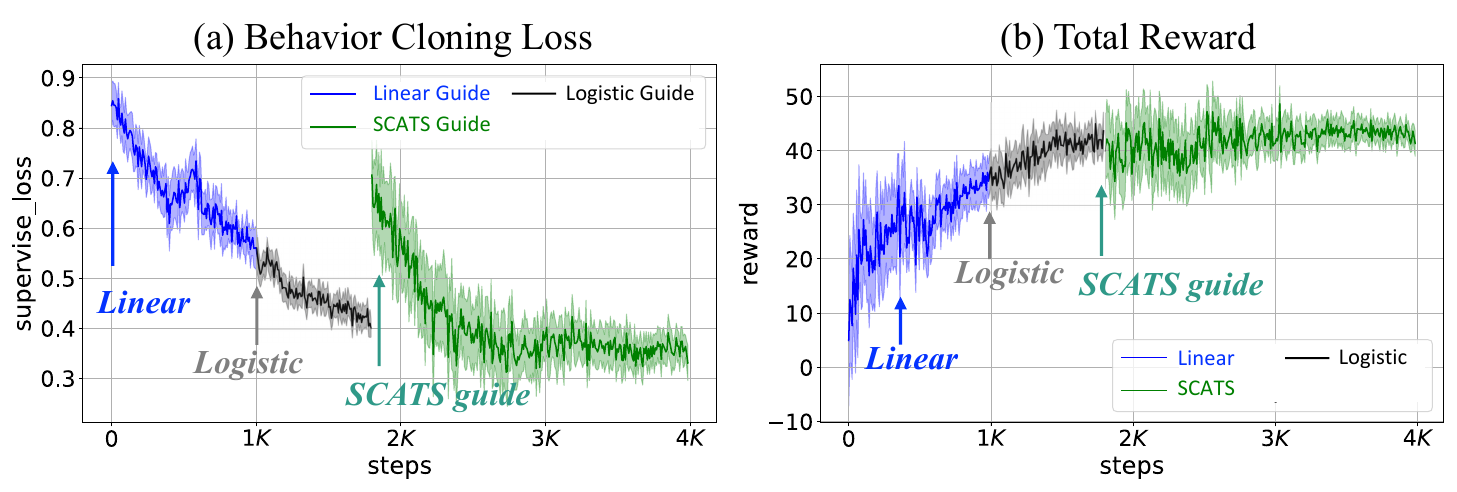}
\caption{Training process under BC: first \textcolor{blue}{linear guide}, then \textcolor{gray}{logistic guide}, and lastly \textcolor{Green}{SCATS guide}. After introducing the linear guide and the logistic guide, the training loss continuously drops. When transitting to the SCATS guide, there was an initial increase in BC loss during the switch, but the reward still had an upward trend throughout the training process. Further, the BC loss steadily decreased. All of these prove the effectiveness of our curriculum learning.}
\label{fig: training}
\end{figure}

\subsubsection{Visualization of Curriculum Training Process} 
\re{Fig. \ref{fig: training} also shows the training process of the curriculum learning. As we can observe, introducing the linear and logistic guides has quite similar and non-distinguishable effects on the training process, and the main reason is that both linear and logistic guides are smooth (in an extreme case, a logistic function with a tiny slope is almost equivalent to a linear function ). Thus, linear and logistic guides constantly help to reduce the BC loss further and increase the reward; when we introduce the advanced SCATS guide, the loss and reward get worse first for a while, mainly due to the SCATS guide being non-smooth, but then, the training gets immediately improved since SCATS is a well-tailored industrial solution for traffic signal control, being configured perfectly for the task.}

\subsubsection{Ablation Studies}
\label{sub:Abluation Studies}

\begin{figure}[t]
\centering
\includegraphics[width=0.95\columnwidth]{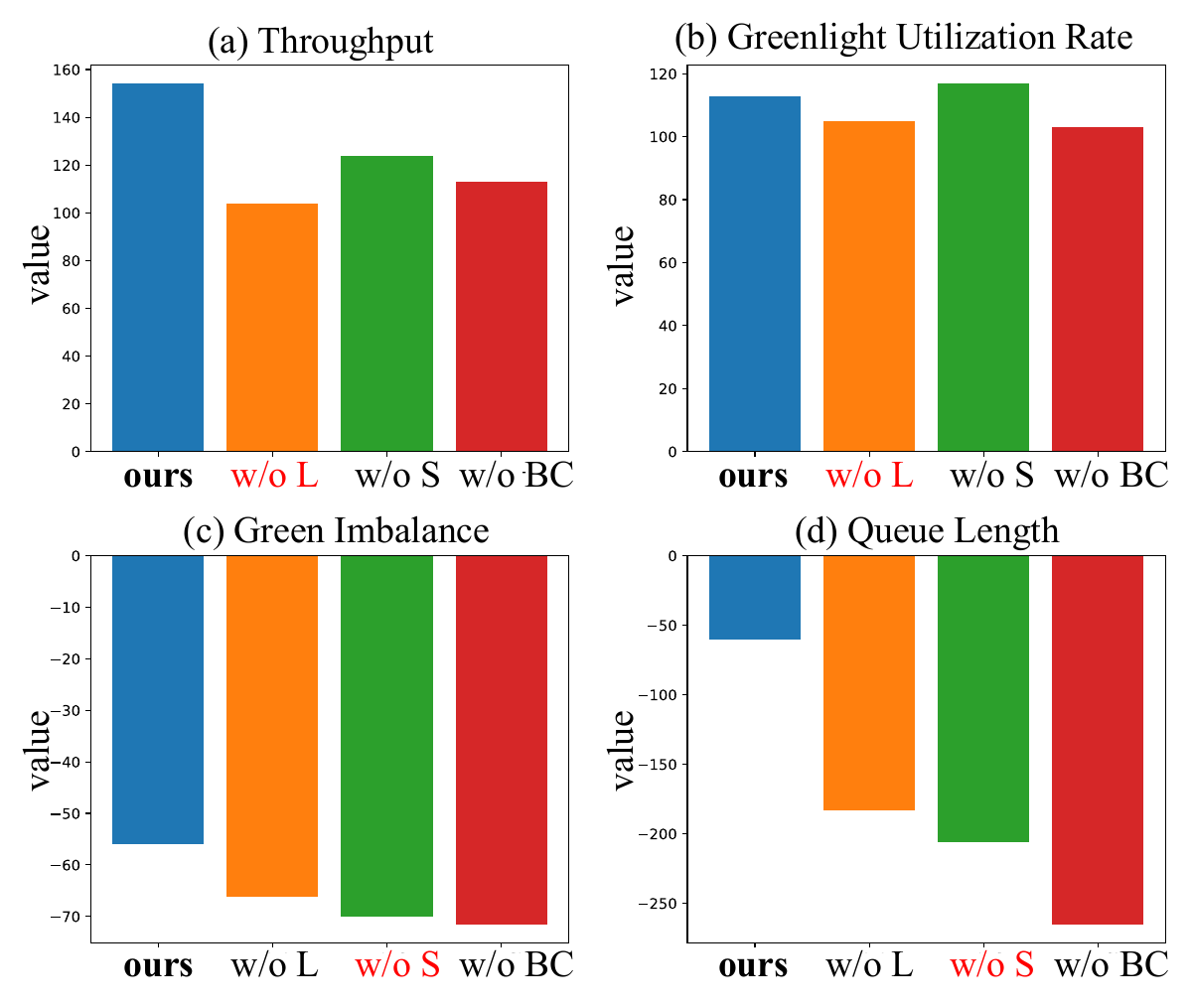}
\caption{Ablation study. Except for the fact that our solution is slightly worse than without BC in Green-light utilization rate, our solution has achieved the best results in another 3 metrics, proving the effectiveness of our design.}
\label{fig: ab}
\end{figure}

Ablation studies are conducted to assess the impact of each component within our GuidedLight. As shown in Fig. \ref{fig: ab}, ``w/o BC'' denotes the model without guidance at all, ``w/o L'' is without guidance from the linear and logistic model (since the two have similar effects, so we do not distinguish them), and ``w/o S'' is without guidance from SCATS.

\re{From Fig. \ref{fig: ab} (a)-(b), the entry-level guide, the linear model, seems to be more critical to achieving higher throughput and green utilization rate. It may be because the linear model follows a straight rule that more cars need longer cycle time, thus allowing more cars to release, resulting in higher throughput and utilization rates. From Fig. \ref{fig: ab} (c)-(d), the advanced guide, SCATS, seems to affect the green imbalance and queue length more: these two indexes are usually more complexly correlated with other traffic factors, and an advanced guide is better. Straightforwardly, the model without any guidance has the worst performance.}


\section{Discussion} 
\label{sec: dis}
\textbf{Choices of Phase Plan:} This paper uses a 4-phase plan as an example, which covers most intersections. It can also handle 3-phase plans by adding masks in state and action. For the 5-phase plan, state and action need to be extended. \textcolor{black}{In addition, due to the integrated features of FRAP~\cite{zheng2019learning}, our method can also handle various types of intersections.}

\textbf{Generalization to Different Intersection Structures:} Given different intersections can have quite different structures, with different numbers of lanes, movements, legs, and shapes (each as T-shape, Y-shape), for a more generalized solution, we recommend using a general plug-in module from GESA \cite{jiang2023a}, which maps various intersections to a unified 4-leg intersection.


\section{CONCLUSIONS}
\label{sec: con}
This paper proposed GuidedLight, an RL-based model for TSC. GuidedLight is delicate to further improve RL methods' applicability in the real world. Several contributions have been made. From the model's perspective: (1) the RL setting is strictly designed following the industry standards in terms of input, output, evaluation, and so on; 
(2) moreover, to pursue the standard cycle-flow relation, we innovatively incorporate traditional yet effective methods such as SCATS as the guidance and use behavior cloning and curriculum learning to teach the agents. From the data's perspective, our simulation scenario is strictly designed to replicate the real-world case. We theoretically prove that the guided RL can guarantee a polynomial sample complexity, and rigid experiments confirm that our GuidedLight could achieve higher performance and uphold industry requirements.

\textbf{Future work}: we will deploy the GuildedLight model in real cities and consider utilizing multi-agent reinforcement learning to facilitate collaborative coordination among multiple intersections, further improving traffic efficiency while ensuring alignment with industry needs.









\bibliographystyle{IEEEtran}
\bibliography{main}

\begin{thebibliography}{10}
\providecommand{\url}[1]{#1}
\csname url@rmstyle\endcsname
\providecommand{\newblock}{\relax}
\providecommand{\bibinfo}[2]{#2}
\providecommand\BIBentrySTDinterwordspacing{\spaceskip=0pt\relax}
\providecommand\BIBentryALTinterwordstretchfactor{4}
\providecommand\BIBentryALTinterwordspacing{\spaceskip=\fontdimen2\font plus
\BIBentryALTinterwordstretchfactor\fontdimen3\font minus \fontdimen4\font\relax}
\providecommand\BIBforeignlanguage[2]{{%
\expandafter\ifx\csname l@#1\endcsname\relax
\typeout{** WARNING: IEEEtran.bst: No hyphenation pattern has been}%
\typeout{** loaded for the language `#1'. Using the pattern for}%
\typeout{** the default language instead.}%
\else
\language=\csname l@#1\endcsname
\fi
#2}}

\bibitem{ketter2023information}
W.~Ketter, K.~Schroer, and K.~Valogianni, ``Information systems research for smart sustainable mobility: A framework and call for action,'' \emph{Information Systems Research}, vol.~34, no.~3, pp. 1045--1065, 2023.

\bibitem{van2016coordinated}
E.~Van~der Pol and F.~A. Oliehoek, ``Coordinated deep reinforcement learners for traffic light control,'' \emph{Proceedings of Learning, Inference and Control of Multi-agent Systems (at NIPS 2016)}, vol.~1, 2016.

\bibitem{mousavi2017traffic}
S.~S. Mousavi, M.~Schukat, and E.~Howley, ``Traffic light control using deep policy-gradient and value-function-based reinforcement learning,'' \emph{IET Intelligent Transport Systems}, vol.~11, no.~7, pp. 417--423, 2017.

\bibitem{wei2018intellilight}
H.~Wei, G.~Zheng, H.~Yao, and Z.~Li, ``Intellilight: A reinforcement learning approach for intelligent traffic light control,'' in \emph{Proceedings of the 24th ACM SIGKDD International Conference on Knowledge Discovery \& Data Mining}, 2018, pp. 2496--2505.

\bibitem{zheng2019learning}
G.~Zheng, Y.~Xiong, X.~Zang, J.~Feng, H.~Wei, H.~Zhang, Y.~Li, K.~Xu, and Z.~Li, ``Learning phase competition for traffic signal control,'' in \emph{Proceedings of the 28th ACM International Conference on Information and Knowledge Management}, 2019, pp. 1963--1972.

\bibitem{jiang2022general}
H.~Jiang, Z.~Li, L.~Bai, R.~Zhao, \emph{et~al.}, ``A general scenario-agnostic reinforcement learning for traffic signal control,'' 2022.

\bibitem{lu2023dualight}
J.~Lu, J.~Ruan, H.~Jiang, Z.~Li, H.~Mao, and R.~Zhao, ``Dualight: Enhancing traffic signal control by leveraging scenario-specific and scenario-shared knowledge,'' \emph{arXiv preprint arXiv:2312.14532}, 2023.

\bibitem{hunt1982scoot}
P.~Hunt, D.~Robertson, R.~Bretherton, and M.~C. Royle, ``The {SCOOT} on-line traffic signal optimisation technique,'' \emph{Traffic Engineering \& Control}, vol.~23, no.~4, 1982.

\bibitem{jiang2023a}
\BIBentryALTinterwordspacing
H.~Jiang, Z.~Li, L.~Bai, Z.~Li, and R.~Zhao, ``A general scenario-agnostic reinforcement learning for traffic signal control,'' 2023. [Online]. Available: \url{https://openreview.net/forum?id=RKMbC8Tslx}
\BIBentrySTDinterwordspacing

\bibitem{lou2022meta}
Y.~Lou, J.~Wu, and Y.~Ran, ``Meta-reinforcement learning for multiple traffic signals control,'' in \emph{Proceedings of the 31st ACM International Conference on Information \& Knowledge Management}, 2022, pp. 4264--4268.

\bibitem{zang2020metalight}
X.~Zang, H.~Yao, G.~Zheng, N.~Xu, K.~Xu, and Z.~Li, ``Metalight: Value-based meta-reinforcement learning for traffic signal control,'' in \emph{Proceedings of the AAAI Conference on Artificial Intelligence}, vol.~34, no.~01, 2020, pp. 1153--1160.

\bibitem{el2013multiagent}
S.~El-Tantawy, B.~Abdulhai, and H.~Abdelgawad, ``Multiagent reinforcement learning for integrated network of adaptive traffic signal controllers ({MARLIN-ATSC}): Methodology and large-scale application on downtown {T}oronto,'' \emph{IEEE Transactions on Intelligent Transportation Systems}, vol.~14, no.~3, pp. 1140--1150, 2013.

\bibitem{wong2023deep}
A.~Wong, T.~B{\"a}ck, A.~V. Kononova, and A.~Plaat, ``Deep multiagent reinforcement learning: Challenges and directions,'' \emph{Artificial Intelligence Review}, vol.~56, no.~6, pp. 5023--5056, 2023.

\bibitem{2019PressLight}
H.~Wei, C.~Chen, G.~Zheng, K.~Wu, V.~Gayah, K.~Xu, and Z.~Li, ``Presslight: Learning max pressure control to coordinate traffic signals in arterial network,'' in \emph{Proceedings of the 25th ACM SIGKDD International Conference on Knowledge Discovery \& Data Mining}, 2019, pp. 1290--1298.

\bibitem{ruan2024coslight}
J.~Ruan, Z.~Li, H.~Wei, H.~Jiang, J.~Lu, X.~Xiong, H.~Mao, and R.~Zhao, ``Coslight: Co-optimizing collaborator selection and decision-making to enhance traffic signal control,'' \emph{arXiv preprint arXiv:2405.17152}, 2024.

\bibitem{jiang2024x}
H.~Jiang, Z.~Li, H.~Wei, X.~Xiong, J.~Ruan, J.~Lu, H.~Mao, and R.~Zhao, ``X-light: Cross-city traffic signal control using transformer on transformer as meta multi-agent reinforcement learner,'' \emph{arXiv preprint arXiv:2404.12090}, 2024.

\bibitem{du2024felight}
X.~Du, Z.~Li, C.~Long, Y.~Xing, S.~Y. Philip, and H.~Chen, ``Felight: Fairness-aware traffic signal control via sample-efficient reinforcement learning,'' \emph{IEEE Transactions on Knowledge and Data Engineering}, 2024.

\bibitem{li2021improved}
Z.~Li, Q.~Zeng, Y.~Liu, J.~Liu, and L.~Li, ``An improved traffic lights recognition algorithm for autonomous driving in complex scenarios,'' \emph{International Journal of Distributed Sensor Networks}, vol.~17, no.~5, p. 15501477211018374, 2021.

\bibitem{manual2000highway}
H.~C. Manual, ``Highway capacity manual,'' \emph{Washington, DC}, vol.~2, no.~1, 2000.

\bibitem{lloyd1967american}
P.~Lloyd, ``American, german and british antecedents to pearl and reed's logistic curve,'' \emph{Population Studies}, vol.~21, no.~2, pp. 99--108, 1967.

\bibitem{lowrie1990scats}
P.~Lowrie, ``{SCATS}, {S}ydney co-ordinated adaptive traffic system: A traffic responsive method of controlling urban traffic,'' 1990.

\bibitem{lillicrap2015continuous}
T.~P. Lillicrap, J.~J. Hunt, A.~Pritzel, N.~Heess, T.~Erez, Y.~Tassa, D.~Silver, and D.~Wierstra, ``Continuous control with deep reinforcement learning,'' \emph{arXiv preprint arXiv:1509.02971}, 2015.

\bibitem{narvekar2020curriculum}
S.~Narvekar, B.~Peng, M.~Leonetti, J.~Sinapov, M.~E. Taylor, and P.~Stone, ``Curriculum learning for reinforcement learning domains: A framework and survey,'' \emph{The Journal of Machine Learning Research}, vol.~21, no.~1, pp. 7382--7431, 2020.

\bibitem{roess2004traffic}
R.~P. Roess, E.~S. Prassas, and W.~R. McShane, \emph{Traffic engineering}.\hskip 1em plus 0.5em minus 0.4em\relax Pearson/Prentice Hall, 2004.

\bibitem{fellendorf1994vissim}
M.~Fellendorf, ``{VISSIM}: A microscopic simulation tool to evaluate actuated signal control including bus priority,'' in \emph{64th Institute of Transportation Engineers Annual Meeting}, vol.~32.\hskip 1em plus 0.5em minus 0.4em\relax Springer, 1994, pp. 1--9.

\bibitem{mirchandani2001real}
P.~Mirchandani and L.~Head, ``A real-time traffic signal control system: architecture, algorithms, and analysis,'' \emph{Transportation Research Part C: Emerging Technologies}, vol.~9, no.~6, pp. 415--432, 2001.

\bibitem{koonce2008traffic}
P.~Koonce and L.~Rodegerdts, ``Traffic signal timing manual.'' United States. Federal Highway Administration, Tech. Rep., 2008.

\bibitem{yau2017survey}
K.-L.~A. Yau, J.~Qadir, H.~L. Khoo, M.~H. Ling, and P.~Komisarczuk, ``A survey on reinforcement learning models and algorithms for traffic signal control,'' \emph{ACM Computing Surveys (CSUR)}, vol.~50, no.~3, pp. 1--38, 2017.

\bibitem{wei2019survey}
H.~Wei, G.~Zheng, V.~Gayah, and Z.~Li, ``A survey on traffic signal control methods,'' \emph{arXiv preprint arXiv:1904.08117}, 2019.

\bibitem{prashanth2011reinforcement}
L.~Prashanth and S.~Bhatnagar, ``Reinforcement learning with average cost for adaptive control of traffic lights at intersections,'' in \emph{2011 14th International IEEE Conference on Intelligent Transportation Systems (ITSC)}.\hskip 1em plus 0.5em minus 0.4em\relax IEEE, 2011, pp. 1640--1645.

\bibitem{xu2013study}
L.-H. Xu, X.-H. Xia, and Q.~Luo, ``The study of reinforcement learning for traffic self-adaptive control under multiagent markov game environment,'' \emph{Mathematical Problems in Engineering}, vol. 2013, 2013.

\bibitem{gao2017adaptive}
J.~Gao, Y.~Shen, J.~Liu, M.~Ito, and N.~Shiratori, ``Adaptive traffic signal control: Deep reinforcement learning algorithm with experience replay and target network,'' \emph{arXiv preprint arXiv:1705.02755}, 2017.

\bibitem{casas2017deep}
N.~Casas, ``Deep deterministic policy gradient for urban traffic light control,'' \emph{arXiv preprint arXiv:1703.09035}, 2017.

\bibitem{xu2021hierarchically}
B.~Xu, Y.~Wang, Z.~Wang, H.~Jia, and Z.~Lu, ``Hierarchically and cooperatively learning traffic signal control,'' in \emph{Proceedings of the AAAI Conference on Artificial Intelligence}, vol.~35, no.~1, 2021, pp. 669--677.

\bibitem{liang2019deep}
X.~Liang, X.~Du, G.~Wang, and Z.~Han, ``A deep reinforcement learning network for traffic light cycle control,'' \emph{IEEE Transactions on Vehicular Technology}, vol.~68, no.~2, pp. 1243--1253, 2019.

\bibitem{hochreiter1997long}
S.~Hochreiter and J.~Schmidhuber, ``Long short-term memory,'' \emph{Neural computation}, vol.~9, no.~8, pp. 1735--1780, 1997.

\bibitem{chang2023learning}
J.~D. Chang, K.~Brantley, R.~Ramamurthy, D.~Misra, and W.~Sun, ``Learning to generate better than your llm,'' \emph{arXiv preprint arXiv:2306.11816}, 2023.

\bibitem{schulman2017proximal}
J.~Schulman, F.~Wolski, P.~Dhariwal, A.~Radford, and O.~Klimov, ``Proximal policy optimization algorithms,'' \emph{arXiv preprint arXiv:1707.06347}, 2017.

\bibitem{koenig1993complexity}
S.~Koenig and R.~G. Simmons, ``Complexity analysis of real-time reinforcement learning,'' in \emph{AAAI}, vol.~93, 1993, pp. 99--105.

\bibitem{agarwal2021theory}
A.~Agarwal, S.~M. Kakade, J.~D. Lee, and G.~Mahajan, ``On the theory of policy gradient methods: Optimality, approximation, and distribution shift,'' \emph{The Journal of Machine Learning Research}, vol.~22, no.~1, pp. 4431--4506, 2021.

\bibitem{tekin2015distributed}
C.~Tekin and M.~Van Der~Schaar, ``Distributed online learning via cooperative contextual bandits,'' \emph{IEEE transactions on signal processing}, vol.~63, no.~14, pp. 3700--3714, 2015.

\bibitem{langford2007epoch}
J.~Langford and T.~Zhang, ``The epoch-greedy algorithm for contextual multi-armed bandits,'' \emph{Advances in neural information processing systems}, vol.~20, no.~1, pp. 96--1, 2007.

\bibitem{simchi2022bypassing}
D.~Simchi-Levi and Y.~Xu, ``Bypassing the monster: A faster and simpler optimal algorithm for contextual bandits under realizability,'' \emph{Mathematics of Operations Research}, vol.~47, no.~3, pp. 1904--1931, 2022.

\bibitem{krishnamurthy2020contextual}
A.~Krishnamurthy, J.~Langford, A.~Slivkins, and C.~Zhang, ``Contextual bandits with continuous actions: Smoothing, zooming, and adapting,'' \emph{The Journal of Machine Learning Research}, vol.~21, no.~1, pp. 5402--5446, 2020.

\bibitem{kakade2002approximately}
S.~Kakade and J.~Langford, ``Approximately optimal approximate reinforcement learning,'' in \emph{Proceedings of the Nineteenth International Conference on Machine Learning}, 2002, pp. 267--274.

\bibitem{uchendu2023jump}
I.~Uchendu, T.~Xiao, Y.~Lu, B.~Zhu, M.~Yan, J.~Simon, M.~Bennice, C.~Fu, C.~Ma, J.~Jiao, \emph{et~al.}, ``Jump-start reinforcement learning,'' in \emph{International Conference on Machine Learning}.\hskip 1em plus 0.5em minus 0.4em\relax PMLR, 2023, pp. 34\,556--34\,583.

\end{thebibliography}

\end{document}